\documentclass[%
 preprint, 
superscriptaddress,
 twocolumn,
 amsmath,amssymb,
 aps, physrev,
]{revtex4-2}

\usepackage{graphicx}
\usepackage{dcolumn}
\usepackage{bm}

\usepackage{hyperref}
\hypersetup{
    colorlinks=true,      
    linkcolor=blue,      
    citecolor=green,  
    urlcolor=cyan}  

\usepackage{etoolbox}
\makeatletter
\patchcmd{\bibliography}{\@ifx}{\@gobbletwo}{}{}
\makeatother

\usepackage{orcidlink}

\begin{document}

\preprint{Physical Review D: \href{https://doi.org/10.1103/sdst-3d6p}{10.1103/sdst-3d6p}}

\title{\textbf{Anisotropic diffusion modeling of cosmic-ray lepton propagation} 
}%

\author{V.D. Borisov\,\orcidlink{0009-0005-2862-7133}}
\email[]{sw.vladislav.b@gmail.com}
\affiliation{Faculty of Physics, M.V. Lomonosov Moscow State University (MSU),
119991 Moscow, Russia}
\affiliation{Skobeltsyn Institute of Nuclear Physics, Lomonosov Moscow State University (MSU),
119991 Moscow, Russia}

\author{I.A. Kudryashov\,\orcidlink{0009-0009-1889-6232}}
\email[]{ilya.kudryashov.85@gmail.com}
\affiliation{Skobeltsyn Institute of Nuclear Physics, Lomonosov Moscow State University (MSU),
119991 Moscow, Russia}



\date{April 2, 2026}

\begin{abstract}
    We analyze DAMPE and H.E.S.S. measurements of the total cosmic-ray electron-positron spectrum, together with the AMS-02 positron fraction, using an anisotropic, spatially varying diffusion framework. The diffusion-tensor components are computed via numerical integration of test-particle trajectories in a prescribed Galactic magnetic-field model. We show that simultaneously accounting for the spatial dependence and anisotropy of the diffusion tensor yields an accurate description of the local electron and positron data up to TeV energies. The inferred injection spectral index, $\gamma=2.169$, is fully consistent with expectations from diffusive shock-acceleration theory. In this approach, the observed spectral softening arises naturally from enhanced energy losses experienced by leptons propagating over larger effective distances along the large-scale magnetic field.

\end{abstract}

\maketitle


\section{\label{sec:level1}Introduction}

A detailed understanding of cosmic-ray (CR) propagation in the Galactic magnetic field is essential for establishing a consistent link between the physical processes operating in the interstellar medium and a wide range of observables, including local particle spectra as well as diffuse gamma-ray and radio emission. For example, studies of Galactic radio emission are widely used to reconstruct the structure of the turbulent magnetic field (see, e.g., Refs.~\cite{Hutschenreuter_2024,Heinrich_2024}), while modeling the local spectra of CR nuclei and their connection to gamma-ray emission up to the PeV energy range has been discussed in Ref.~\cite{Prevotat_2024}. Such connections have been extensively investigated using numerical propagation codes such as GALPROP, DRAGON, and PICARD~\cite{StrongMoskalenkoReimer2004,Gaggero2015DRAGON,Kissmann2014PICARD}. The reliability of these models depends both on the realism of the adopted Galactic geometry and the distributions of sources, interstellar gas, and radiation fields, as well as on the choice of the underlying transport framework (see Refs.~\cite{Ackermann2012Diffuse,OrlandoStrong2013} for reviews).

Existing approaches span a wide range of complexity. Analytic Green’s-function solutions~\cite{Webb_1994}, applicable to homogeneous or symmetric media, offer convenient estimates and enable controlled studies of individual sources. However, they require diffusion parameters borrowed from more complete numerical models and become less applicable when describing global diffuse emission; see, e.g., Ref.~\cite{Hanasz_2021} for a review. More advanced diffusion and stochastic codes—such as GALPROP~\cite{Strong1998}, DRAGON2~\cite{Evoli_2017}, and CRPropa~\cite{Batista_2016}—solve the CR transport equation within a three-dimensional Galactic structure and reproduce a broad range of observational data. Nevertheless, the diffusion approximation itself relies on several nontrivial assumptions: a clear separation of spatial scales between the Larmor radius and large-scale magnetic-field gradients, quasilinear interactions with turbulence, and approximate statistical homogeneity on diffusion scales. These assumptions may break down near sources~\cite{Giacinti_2012}, in shock environments, in regions of strong magnetic-field anisotropy, and in the transition between ballistic and diffusive regimes; see, for example, Ref.~\cite{Prosekin_2015}. In contrast, direct numerical methods—such as test-particle trajectory integration in prescribed 3D magnetic fields~\cite{kuhlen_2025} or fully self-consistent magnetohydrodynamics simulations--circumvent the diffusion approximation and explicitly track particle dynamics~\cite{Wibking_2023}. However, the computational cost increases sharply toward lower energies because of the decreasing Larmor radius and the need to resolve an increasingly large number of gyroperiods. As a consequence, direct methods remain practically limited to high-energy particles or to localized regions~\cite{Giacinti_2015}. Diffusion models therefore continue to serve as the primary tool for studying CR transport on Galactic scales, providing a practical balance between physical fidelity and computational feasibility, while direct numerical methods are used to validate their applicability and explore regimes where the diffusion approximation fails.

Over time, diffusion codes have grown substantially more sophisticated: they now incorporate detailed three-dimensional gas maps, modern nuclear-interaction models, realistic interstellar radiation fields, and dedicated synchrotron modules. This progress has improved the realism of simulations and agreement with observations. Nevertheless, many traditional models still employ simplified, often isotropic diffusion prescriptions, assuming that the diffusion coefficient is independent of the magnetic-field orientation or adopting only simple spatial variations. At the same time, reconstructions of the large-scale Galactic magnetic field—from WMAP and Planck data, Faraday-rotation measures, and polarized synchrotron maps—indicate that the regular field component can be comparable in strength to the turbulent one~\cite{Jansson_2012,Unger_2024}. Under such conditions, significantly different transport rates parallel and perpendicular to the magnetic field are physically expected. Numerical and theoretical studies confirm that anisotropic diffusion—manifested in the distinction between $D_{\parallel}$ and $D_{\perp}$—can substantially modify CR propagation, residence times, densities, and spectra throughout the Galaxy. Anisotropic and spatially varying diffusion coefficients have been extensively investigated in works based on direct test-particle integration in turbulent fields (See Refs.~\cite{Casse_2002,Dundovic_2020} and, for extremely small ratios of gyroradius to correlation length, $R_g/L_{\mathrm{cor}}$, Ref.~\cite{kuhlen_2025}), as well as in the framework of nonlinear transport theory~\cite{Shalchi_2009}. These studies show that the ratio $D_{\parallel}/D_{\perp}$ can reach values of order $10^3$, while the magnetic-field structure itself produces substantial spatial inhomogeneity in the diffusion tensor. Consequently, several modified diffusion frameworks—including implementations within DRAGON~\cite{DeLaTorreLuque_2025,ALZetoun2018Anisotropic}—have introduced distinct parallel and perpendicular diffusion coefficients, although simplified magnetic-field geometries (e.g., axisymmetric prescriptions) are still commonly employed.

We argue that a fully anisotropic, tensorial diffusion framework—consistent with a realistic three-dimensional model of the Galactic magnetic field, including both regular and turbulent components—is essential for a proper interpretation of modern experimental data, including direct measurements of CR fluxes and diffuse gamma-ray and radio emission. This is important because diffusion anisotropy can modify local CR spectra and the spatial distribution of emission, thereby impacting conclusions regarding acceleration mechanisms, source properties, and the structure of the interstellar medium. In most existing diffusion codes, the diffusion tensor is implemented in a simplified form. Typically, either only the flux term, $j = -D \nabla \psi$, is approximated or parallel and perpendicular components, $D_{\parallel}$ and $D_{\perp}$, are introduced under assumptions of simple magnetic-field geometry. Such treatments neglect the spatial dependence of the tensor components themselves, as well as derivative contributions that appear in the transport operator $\nabla \cdot (D \nabla \psi)$. In realistic three-dimensional magnetic-field configurations—representative of the actual Galactic structure or its local environments—a global description of diffusion using only two coefficients, $D_{\parallel}$ and $D_{\perp}$, becomes insufficient. Although one can always define locally parallel and perpendicular components in an orthonormal basis aligned with the magnetic field, spatial variations of the regular field (spiral, toroidal, and vertical components) and of the turbulence amplitude lead to significant changes in the local orientation of $\hat{B}$ and in the relative importance of $D_{\parallel}$ and $D_{\perp}$. As a result, the diffusion tensor in any fixed global coordinate system becomes highly inhomogeneous and anisotropic, with its components reflecting both the large-scale magnetic-field structure and the local turbulence spectrum and correlation length. A realistic diffusion model must therefore determine $D(\mathbf{r})$ not merely as a free parameter tuned to observations but as a quantity calibrated against direct numerical calculations of diffusion coefficients in a prescribed turbulent field. This approach ensures consistency between macroscopic diffusion models and microscopic test-particle trajectory integrations, linking first-principles estimates of diffusion coefficients with the computational efficiency of large-scale transport simulations.

Despite substantial progress in developing three-dimensional diffusion models and improving reconstructions of the Galactic magnetic field, the extent to which transport anisotropy affects observed CR spectra remains an open question. Cosmic-ray electrons are particularly sensitive to transport details because their lifetimes are dominated by strong radiative losses and decrease rapidly with energy. In this regime, both the local spectrum and the spatial distribution of electrons serve as direct diagnostics of how realistically a model captures particle transport in a three-dimensional, intrinsically anisotropic environment.

Modern measurements by DAMPE~\cite{dampe2017}, CALET~\cite{adriani2017}, Fermi-LAT~\cite{fermilat2017}, H.E.S.S.~\cite{hess2024}, and AMS-02~\cite{ams02electrons2014} indicate that the electron spectrum follows a power law with index $\gamma \approx 3$ up to several hundred GeV. Recent measurements of local electrons also suggest a small change of slope around $E \sim 40$~GeV~\cite{EvoliBlasiAmatoAloisio2020}. At energies $E \gtrsim 1$~TeV, the spectrum exhibits a significant softening~\cite{hess2024}, corresponding to an effective index $\gamma \approx 4$. The spectral break near 1~TeV is usually attributed to a combination of effects: the limited lifetime of high-energy electrons, details of the acceleration processes in sources, and the decreasing number of nearby young sources at $E \gtrsim 1$~TeV, consistent with the shrinking lepton propagation horizon~\cite{thoudam2024}. Despite notable progress, no single model currently reproduces the electron spectrum across the full energy range without considerable parameter tuning. It is generally assumed that electron acceleration proceeds via diffusive shock acceleration, producing an initial power-law spectrum with index $\gamma$. During propagation, electrons lose energy predominantly through inverse Compton scattering and synchrotron radiation; ionization losses become negligible above a few tens of GeV. Depending on the relative magnitudes of the characteristic diffusion and loss times, three regimes may occur:

\emph{Loss-dominated regime.} Particles lose energy more rapidly than they diffuse, resulting in a spectral steepening by unity.

\emph{Diffusion-dominated regime.} Particles escape the region before losing significant energy; losses do not modify the spectrum.

\emph{Mixed regime.} Diffusion and loss times are comparable; the spectral steepening assumes an intermediate value $\xi \in [0,1]$.

Estimates of the diffusion coefficient inferred from secondary-to-primary nuclei ratios~\cite{dellatorreluque2021,niu2018}, together with direct numerical calculations of the diffusion tensor~\cite{kuhlen_2025}, yield characteristic values $D \sim 10^{28}$--$10^{30}$~cm$^2$\,s$^{-1}$ for energies ranging from tens of GeV to tens of TeV. These values correspond to a mixed regime transitioning into a loss-dominated regime, with the transition energy depending on the specific diffusion model. Additional spectral modifications arise from the energy dependence of diffusion, $D \propto E^{\delta}$, where $\delta \sim 1/3$--$1/2$ corresponds to the turbulence spectrum of the interstellar magnetic field (Kolmogorov and Kraichnan limits).

The anisotropy of the Galactic magnetic field is one of the key factors determining the transport of relativistic electrons. The regular component—characterized by a predominantly spiral structure in the disk—sets the principal direction of diffusion, whereas the turbulent component, shaped by the inhomogeneous distribution of sources and irregularities of the interstellar medium, controls the scattering rates and the turbulence spectrum. Under these conditions, diffusion becomes strongly anisotropic: the parallel coefficient $D_{\parallel}$ often exceeds the perpendicular coefficient $D_{\perp}$ by several orders of magnitude. Consequently, the local electron spectrum may appear softened or hardened depending on the magnetic-field configuration, source gradients, and diffusion-tensor geometry~\cite{borisov2025,borisov2025_2}, and is typically expressed as $E^{-\Gamma}$, where $\Gamma = \gamma + \delta + \xi$.

This paper is organized as follows. Section~\ref{sec:equation} describes the anisotropic diffusion model with energy losses employed in this work, along with the main features of our discretization of the diffusion equation. The construction of the diffusion tensor is discussed in detail in Sec.~\ref{sec:tensor}. Section~\ref{sec:loss} presents the software module responsible for computing energy losses in the anisotropic diffusion framework. In Sec.~\ref{sec:results}, we compare different scenarios for describing the electron spectrum and its spatial distribution under various levels of approximation to the diffusion equation. We conclude with a discussion of our results in Sec.~\ref{sec:conclusion}.

\section{\label{sec:equation}Discretization of the Anisotropic Diffusion Operator with Energy Losses}

The propagation of relativistic electrons and positrons in the interstellar medium is governed by the transport equation,
\begin{equation}
\label{eq:transport}
\begin{split}
\frac{\partial\psi(\mathbf{x},E)}{\partial t}
&= \nabla_i \!\left( D_{ij}(\mathbf{x},E)\, \nabla_j \psi(\mathbf{x},E) \right) \\
 &- \frac{\partial}{\partial E}\!\left[\, \alpha(\mathbf{x},E)\, \psi(\mathbf{x},E) \right] 
 + Q(\mathbf{x},E),
 \end{split}
\end{equation}
where $\psi(\mathbf{x},E)$ is the phase space density, 
$D_{ij}(\mathbf{x},E)$ is the anisotropic spatial diffusion tensor,  
$\alpha(\mathbf{x},E)$ describes continuous energy losses, and 
$Q(\mathbf{x},E)$ is the source term with a prescribed injection spectrum.  
Throughout this work, we consider the stationary form of Eq.~(\ref{eq:transport}), which is appropriate for describing the Galactic background lepton population because the characteristic diffusion time $\tau_{\rm diff}$ and the energy-loss time $\tau_{\rm loss}$ are both short compared to the evolutionary timescales of the large-scale Galactic magnetic field and the gas distribution.

The tensor $D_{ij}(\mathbf{x},E)$ is spatially dependent and generally off-diagonal due to the presence of an ordered Galactic magnetic field. Its energy dependence follows from the rigidity scaling of the parallel and perpendicular diffusion coefficients.  
A key feature of our approach is that the components of $D_{ij}(\mathbf{x},E)$ are computed directly within a physically small volume, where diffusion along and across the local mean magnetic field can be evaluated by integrating test-particle trajectories for a specified turbulence level. The resulting local diffusion tensor is then rotated into the global coordinate system associated with the large-scale Galactic magnetic field.  
This procedure allows the use of an arbitrary (not necessarily axisymmetric) configuration of the regular magnetic field, in contrast to most diffusion models, and therefore resolves local magnetic inhomogeneities with improved accuracy.

In this section, we present a finite-difference discretization of the operator on a four-dimensional grid $(x,y,z,p)$. The scheme preserves the full tensorial structure of the anisotropic diffusion operator, retains the correct transport properties along and across magnetic-field lines, and ensures numerical stability in the presence of strong radiative energy losses.

We denote the spatial grid spacings by $(h_x, h_y, h_z)$ and employ a logarithmic momentum grid, which is a natural choice for leptonic transport where the dominant losses scale as $|\dot{p}| \propto p^2$.

\emph{Diagonal diffusion terms.} For the diagonal components $D_{ii}$ we employ a face-centered discretization
that preserves the conservative structure of $\nabla\!\cdot(D\nabla\psi)$.
The coefficient at the interface $i+\tfrac12$ is approximated as
\begin{equation}
D_{ii}^{\,i+\frac12}
= \frac12\!\left[ D_{ii}(\mathbf{x}) + D_{ii}(\mathbf{x}+\hat{i}) \right],
\end{equation}
leading to the discrete contribution
\begin{align}
\label{eq:diag}
\mathcal{L}_{ii}[\psi] &\approx
\frac{D_{ii}^{\,i+1/2}}{h_i^2}\, \psi_{\mathbf{x}+\hat{i}}
+ \frac{D_{ii}^{\,i-1/2}}{h_i^2}\, \psi_{\mathbf{x}-\hat{i}}
\nonumber\\
&\quad
-\left(\frac{D_{ii}^{\,i+1/2}+D_{ii}^{\,i-1/2}}{h_i^2}\right)\psi_{\mathbf{x}} .
\end{align}
This symmetric stencil guarantees second-order accuracy for smooth $D_{ii}$,
preserves the self-adjoint, negative-definite form of the diffusion operator,
and therefore ensures numerical stability together with the physically correct
damping of small-scale modes. In the limit of spatially uniform diffusion, the
scheme reduces to the standard seven-point Laplacian in three dimensions.

\emph{Cross-terms.} In the presence of a large-scale magnetic field $\mathbf{B}$,
particle transport is anisotropic
\begin{equation}
\begin{split}
D_{ij} = D_{\parallel} b_i b_j
        + D_{\perp}\left(\delta_{ij}-b_i b_j\right),\\
\qquad
b_i = B_i/|\mathbf{B}|.
\end{split}
\end{equation}
In Cartesian coordinates this produces nonzero off-diagonal
components $D_{ij}$ for $i\neq j$, corresponding to correlated diffusion
in different spatial directions.

The mixed derivative term
\begin{equation}
\partial_i(D_{ij}\partial_j\psi)
\end{equation}
is discretized using a symmetric four-point corner stencil
\begin{equation}
\partial_i\partial_j\psi
\;\approx\;
\frac{
\psi_{\hat{i}+\hat{j}}
- \psi_{\hat{i}-\hat{j}}
- \psi_{-\hat{i}+\hat{j}}
+ \psi_{-\hat{i}-\hat{j}}
}{
4 h_i h_j}.
\end{equation}
To preserve self-adjoint-ness of the full operator, the diffusion tensor is locally symmetrized
\begin{equation}
D^{\mathrm{sym}}_{ij}
= \frac12\big( D_{ij} + D_{ji} \big),
\end{equation}
and each of the four diagonal neighbors contributes
\begin{equation}
\pm\, \frac{D_{ij}^{\mathrm{sym}}}{4 h_i h_j}.
\end{equation}
This discretization reproduces anisotropic transport along curved magnetic-field
lines, avoids spurious grid-alignment artifacts, and remains second-order
accurate for smooth $D_{ij}(\mathbf{x})$.

\emph{Terms involving spatial gradients of $D_{ij}$.} In realistic Galactic magnetic-field and turbulence models,
the diffusion tensor varies substantially across the Galaxy.
The divergence operator therefore contains additional contributions
\begin{equation}
(\partial_i D_{ij})\, \partial_j\psi ,
\end{equation}
which must be included explicitly whenever $D_{ij}(\mathbf{x})$
has strong spatial gradients.

The spatial derivative of the tensor is approximated via a centered difference
\begin{equation}
(\partial_i D_{ij})_c
  \approx \frac{ D_{ij}(\mathbf{x}+\hat{i}) - D_{ij}(\mathbf{x}-\hat{i}) }{2 h_i},
\end{equation}
and the derivative of $\psi$ analogously
\begin{equation}
(\partial_j\psi)_c
  \approx \frac{ \psi_{\mathbf{x}+\hat{j}} - \psi_{\mathbf{x}-\hat{j}} }{2 h_j }.
\end{equation}
The product contributes to the $\pm\hat{j}$ neighbors with coefficients
\begin{equation}
\psi_{\mathbf{x}\pm\hat{j}}
\;\approx\;
\pm\,
\frac{ D_{ij}(\mathbf{x}+\hat{i}) - D_{ij}(\mathbf{x}-\hat{i}) }
     {4 h_i h_j}.
\end{equation}

These terms are necessary to ensure that the full discrete operator correctly approximates $\nabla_i(D_{ij}\nabla_j\psi)$ in the presence of
spatially varying anisotropic diffusion, especially where magnetic-field
gradients or turbulence transitions are sharp.

\emph{Momentum-space drift from energy losses.} Continuous energy losses---dominated by synchrotron radiation,
inverse Compton scattering, bremsstrahlung and ionization---enter the transport equation
as a convective (drift) term in momentum space.  
On a logarithmic momentum grid, the cell spacing is
\begin{equation}
\Delta p_j = h_p \ln 10 \, E_j,
\end{equation}
where $h_p$ is the step size in the logarithm of energy and $E_j$ is the energy at node $j$, and the face-centered loss coefficients are evaluated as
\begin{equation}
\label{eq:loss1}
\alpha_{j\pm\frac12}
= \frac12\left[ \alpha(\mathbf{x},p_j) + \alpha(\mathbf{x},p_{j\pm1}) \right].
\end{equation}
The corresponding upwinded discrete operator becomes
\begin{equation}
\label{eq:loss2}
\mathcal{L}_p[\psi]
  = \frac{\alpha_{j+\frac12}}{\Delta p_j}\, \psi_{j+1}
  - \frac{\alpha_{j-\frac12}}{\Delta p_j}\, \psi_{j}.
\end{equation}

Because physical losses satisfy $\alpha>0$, the drift is always directed toward
lower momenta; hence an upwind discretization is required to avoid unphysical
oscillatory solutions and to maintain monotonicity of $\psi$.
At high energies ($\gtrsim$ few $\times 100\,$GeV) this term dominates
over spatial diffusion and defines the effective cooling horizon for leptons.

\section{\label{sec:tensor}Calculation of the Diffusion Tensor in the Presence of Regular Magnetic Field}

In a sufficiently small volume (i.e., on scales comparable to the displacements required for particles to reach the diffusive regime), a local coordinate system $(x',y',z')$ can be introduced such that the $z'$ axis is aligned with the regular magnetic field
$\mathbf{B}_0 = (B_x, B_y, B_z)$.  
In this local field-aligned frame, the diffusion tensor is diagonal,
\begin{equation}
\mathbf{D}' =
\begin{pmatrix}
D_{\perp} & 0 & 0 \\
0 & D_{\perp} & 0 \\
0 & 0 & D_{\parallel}
\end{pmatrix},
\label{eq:Dloc}
\end{equation}
where $D_{\parallel}$ and $D_{\perp}$ denote the parallel and perpendicular diffusion coefficients obtained from test-particle simulations in a prescribed turbulent field.

To express the diffusion tensor in the global Galactic Cartesian coordinates $(x,y,z)$, associated with the large-scale structure of the Galaxy, we perform a rotation using the matrix $\mathbf{R}$:
\begin{equation}
\mathbf{D}^{\rm fld} = \mathbf{R}\, \mathbf{D}' \, \mathbf{R}^{T}.
\label{eq:Drot}
\end{equation}

The orientation of the regular magnetic field $\mathbf{B}_0$ is parametrized by the polar and azimuthal angles
\begin{equation}
\begin{split}
&\cos\Theta = \frac{B_z}{|\mathbf{B}_0|}, \qquad
\sin\Theta = \frac{\sqrt{B_x^2 + B_y^2}}{|\mathbf{B}_0|}, \\
&\cos\Phi = \frac{B_x}{\sqrt{B_x^2 + B_y^2}}, \qquad
\sin\Phi = \frac{B_y}{\sqrt{B_x^2 + B_y^2}}.
\end{split}
\end{equation}whenever $B_x^2 + B_y^2 \neq 0$; otherwise, the azimuthal angle is undefined and the rotation reduces to alignment with the $z$ axis. The corresponding rotation matrix is
\begin{equation}
\mathbf{R} =
\begin{pmatrix}
\cos\Phi \cos\Theta & \sin\Phi & \cos\Phi \sin\Theta \\
-\sin\Phi \cos\Theta & \cos\Phi & -\sin\Phi \sin\Theta \\
-\sin\Theta & 0 & \cos\Theta
\end{pmatrix}.
\label{eq:rot}
\end{equation}

The regular field $\mathbf{B}_0$ is adopted from the Unger24 model~\cite{Unger_2024}, which refines standard spiral-arm based descriptions of the large-scale Galactic field. The amplitude of the turbulent component $\delta B(x,y,z)$ is modeled analytically using the JF12 prescription~\cite{Jansson_2012}.

We assume isotropic turbulence with a constant correlation length $\ell_{\mathrm{c}} = 50$ pc, following Ref.~\cite{kuhlen_2025}. The correlation length is not precisely known, and variations in its value affect the characteristic energy at which the transition occurs from the diffusion of charged particles along magnetic field lines to the diffusion of the field lines themselves. This effect becomes significant only at energies of a few PeV and has negligible impact within the energy range considered here ($E \lesssim 100$ TeV).
This approximation enables a physically consistent calculation of the parallel and perpendicular diffusion coefficients as functions of the local regular magnetic field without introducing additional model-dependent uncertainty.  
Although more complex scenarios are often discussed in the literature (including enhanced turbulence near supernova remnants or in the Galactic center region~\cite{Iffrig2017}, as well as anisotropic turbulence), the present isotropic approximation is sufficient to capture the dominant effects of anisotropic diffusion and avoids introducing poorly constrained parameters for which no comprehensive numerical benchmarks currently exist.

The parallel and perpendicular diffusion coefficients are computed using test-particle trajectory integration in the combined regular and turbulent magnetic field, following the method described in Ref.~\cite{borisov2025}. In practice, large ensembles of particles are propagated through the prescribed magnetic field, and their mean squared displacements along and across the local magnetic field are recorded. These displacements, evaluated as functions of energy and position, directly yield the local parallel and perpendicular components of the diffusion tensor. This trajectory-based approach captures the anisotropic nature of cosmic-ray transport, including the reduced perpendicular diffusion at low energies when particle gyroradii are smaller than the turbulence correlation length, and the transition toward more isotropic diffusion at higher energies.

Once $D_{\parallel}$ and $D_{\perp}$ are determined locally in the field-aligned frame, the full diffusion tensor in Galactic coordinates is obtained by applying the rotation (\ref{eq:Drot}). In general, the resulting diffusion tensor is symmetric and contains up to six independent components. Its spatial variation reflects the nonuniform redistribution of cosmic rays driven by the changing magnetic-field geometry and turbulence properties throughout the Galactic volume.

\section{\label{sec:loss}Energy Losses of Relativistic Electrons and Positrons}

We solve the stationary diffusion-loss transport equation for cosmic-ray electrons and positrons, we implement a numerical module that evaluates the total energy-loss function 
$\alpha(\mathbf{x},E)$ at each spatial point and energy node.  
The loss operator enters the transport equation in the standard form
\begin{equation}
\frac{\partial}{\partial E}\!\left[\alpha(\mathbf{x},E)\, \psi(\mathbf{x},E)\right]
\end{equation}
and is discretized using an upwind finite-difference scheme Eqs.~\eqref{eq:loss1} and \eqref{eq:loss2}.

The total energy-loss rate is decomposed into the four dominant channels,
\begin{equation}
\alpha(\mathbf{x},E)
= b_{\rm syn} + b_{\rm IC} + b_{\rm brem} + b_{\rm coul},
\end{equation}
each evaluated locally using the magnetic field, interstellar radiation field (ISRF) energy densities, gas density, and thermal electron density.

The magnetic field affects the transport equation through the synchrotron loss term via the local magnetic energy density
$U_B = B^2/(8\pi)$.  
Throughout this work we employ the same large-scale Galactic magnetic-field model as used in our computation of the diffusion tensor, including the disk, toroidal-halo, and X-shaped poloidal components.

The ISRF is modeled as a sum of the cosmic microwave background (CMB) and IR, optical, and UV components.
For the non-CMB components, we adopt an exponential spatial distribution,

\begin{equation}
U_{\rm ph}(R,z) =
U_0
\exp\!\left[-\frac{R-R_0}{R_{\rm ph}}\right]
\exp\!\left(-\frac{|z|}{z_{\rm ph}}\right),
\end{equation}

where $R=\sqrt{x^2+y^2}$ is the Galactocentric radius and
$R_0 = 8.2\,{\rm kpc}$ is the solar Galactocentric distance.
The CMB component is assumed to be spatially uniform.

Inverse Compton (IC) losses are computed using the computationally efficient approximation of the Klein-Nishina cross section
proposed in Ref.~\cite{Moderski2005ApJ627}.
In this approach, each radiation component is characterized by an
effective photon energy,

\begin{equation}
\epsilon_{\rm eff} = 2.7\,k_{\rm B}T ,
\end{equation}

corresponding to a blackbody radiation field with temperature $T$.

The interstellar gas distribution is modeled as an exponential disk

\begin{equation}
n_{\rm gas}(R,z) =
n_0
\exp\!\left[-\frac{R-R_0}{R_n}\right]
\exp\!\left(-\frac{|z|}{z_n}\right),
\end{equation}

while the thermal electron density is assumed to decrease exponentially
with height above the Galactic plane,

\begin{equation}
n_e(z) = n_{e,0}\exp\!\left(-\frac{|z|}{z_e}\right).
\end{equation}

The parameters of the radiation and gas distributions used in the
calculation of electron energy losses are summarized in
Table~\ref{tab:isrf_parameters}.

\begin{table}[b]
\caption{\label{tab:isrf_parameters}
Parameters of the interstellar radiation field and local interstellar medium
used for the calculation of electron energy losses.}
\begin{ruledtabular}
\begin{tabular}{cccc}
Component & $U_0$ (eV cm$^{-3}$) & $T$ (K) & $\epsilon_{\rm eff}$ (eV) \\
\hline
CMB      & 0.26 & 2.725  & $6.3\times10^{-4}$ \\
IR       & 0.50 & 30     & $7.0\times10^{-3}$ \\
Optical  & 0.50 & 5000   & 1.16 \\
UV       & 0.10 & 20000  & 4.65 \\
\hline
\multicolumn{4}{c}{Local ISM parameters} \\
$n_0$    & \multicolumn{3}{c}{$1\ {\rm cm^{-3}}$} \\
$n_{e,0}$& \multicolumn{3}{c}{$0.03\ {\rm cm^{-3}}$} \\
$R_{\rm ph}$ & \multicolumn{3}{c}{$5\ {\rm kpc}$} \\
$z_{\rm ph}$ & \multicolumn{3}{c}{$1\ {\rm kpc}$} \\
$R_n$ & \multicolumn{3}{c}{$4\ {\rm kpc}$} \\
$z_n$ & \multicolumn{3}{c}{$0.2\ {\rm kpc}$} \\
$z_e$ & \multicolumn{3}{c}{$1\ {\rm kpc}$} \\
\end{tabular}
\end{ruledtabular}
\end{table}

\begin{figure}[h]
\includegraphics[width=0.5\textwidth]{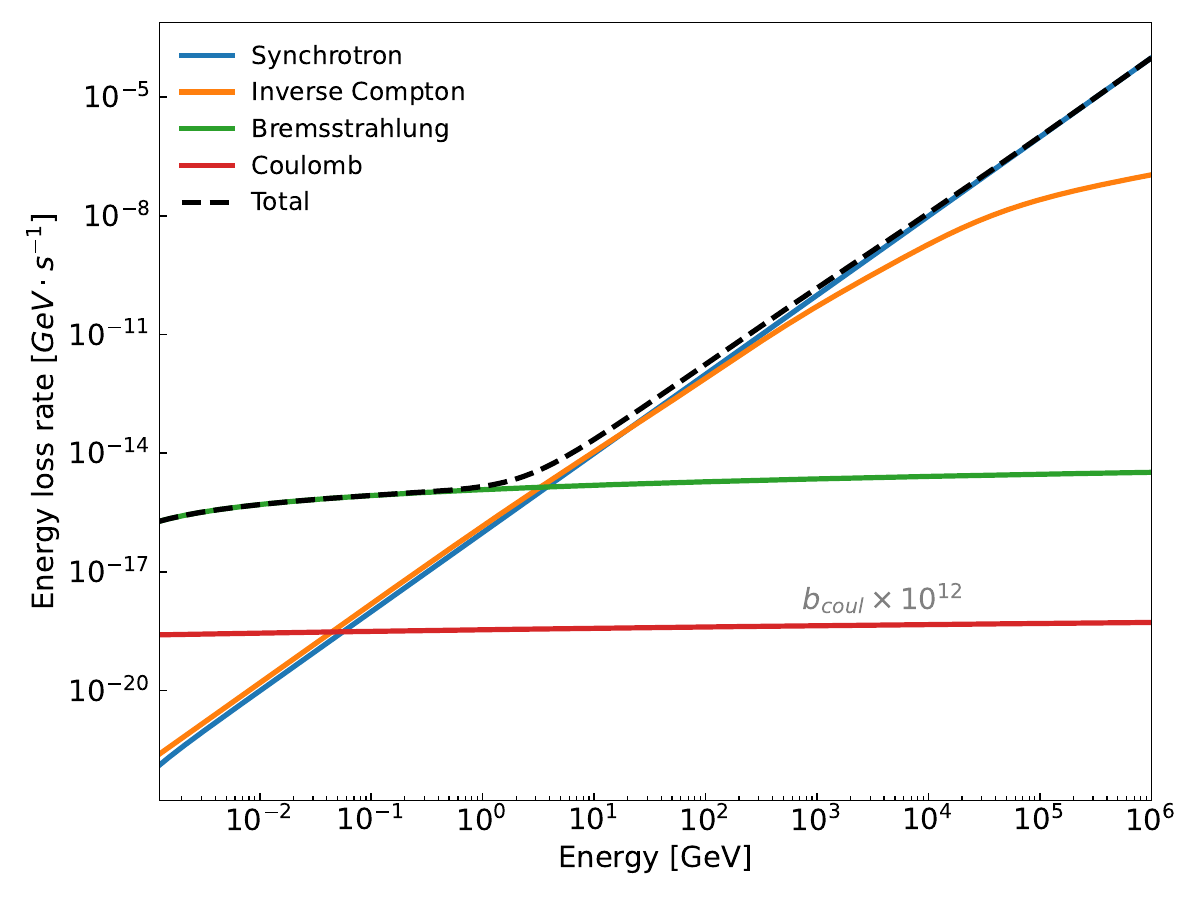}
\caption{\label{fig:loss} Energy-loss rate as a function of lepton energy at the Solar position. The magnetic field is taken from the model~\cite{Unger_2024}, ISRF energy densities from Ref.~\cite{Delahaye2010}, and radiative and Coulomb loss formulas from Ref.~\cite{BlumenthalGould1970}.}
\end{figure}

Figure~\ref{fig:loss} shows the energy dependence of the individual loss terms at the Solar position, $(x,y,z)=(-8.2,0,0.2)\,{\rm kpc}$, in a Cartesian coordinate system centered on the Galactic center.

Given the local magnetic energy density $U_B$, the synchrotron cooling rate is
\begin{equation}
b_{\rm syn}(E)
= \frac{4}{3}\,\sigma_T c\, \gamma^2\, U_B,
\end{equation}
where $\gamma = E/(m_e c^2)$.  
Both ordered and turbulent magnetic-field components are included.

The inverse Compton loss rate is obtained by summing the contributions from all ISRF components,
\begin{equation}
b_{\rm IC}(E)
= \sum_i \frac{4}{3}\,\sigma_T c\,\gamma^2\, U_{{\rm ph},i}\,
F_{\rm KN}(E,T_i),
\end{equation}
where $F_{\rm KN}$ accounts for Klein-Nishina suppression.  
For the suppression factor, we use the approximation
\begin{equation}
F_{\rm KN}(q) = (1 + 4q)^{-3/2},
\qquad
q = \frac{4\gamma\,\epsilon_{\rm eff}}{m_e c^2}.
\end{equation}

Bremsstrahlung losses depend on the local gas density $n_{\rm gas}$.  
In the relativistic regime, we employ the standard approximation
\begin{equation} 
\begin{split} 
b_{\rm brem}(E) &\simeq 1.51\times10^{-16}\, n_{\rm gas}\ \times\\ &\bigl[\ln(\gamma) + 0.36\bigr] \quad {\rm GeV\,s^{-1}}, \end{split} 
\end{equation}
which provides an accurate description across the relevant energy range.

Coulomb losses depend primarily on the thermal electron density $n_e$:
\begin{equation} 
\begin{split} 
b_{\rm coul}(E) &\simeq 3.3\times10^{-29}\, n_e \times \\ &\left[ \frac{\ln(\gamma/n_e)}{75} + 0.2 \right] \quad {\rm GeV\,s^{-1}}. \end{split} 
\end{equation}

This form captures the characteristic logarithmic dependence of low-energy interactions with the ambient plasma. At energies above several tens of GeV, synchrotron and inverse Compton losses dominate, whereas bremsstrahlung and Coulomb losses become subdominant. The present treatment reproduces the onset of the Klein-Nishina regime for scattering on the UV radiation field at energies $\gtrsim 40$~GeV.

The resulting synchrotron and IC cooling rates are consistent with those adopted in the Unger24 framework and with the loss model of Ref.~\cite{EvoliBlasiAmatoAloisio2020}. They are somewhat larger than the corresponding rates used in GALPROP, reflecting the stronger magnetic field and higher radiation-field energy densities assumed in the present model. Overall, the resulting cooling rate provides a consistent description of the local electron spectrum within the anisotropic diffusion framework.

\section{\label{sec:results}Results}
We consider a diffusion model in the mixed transport regime, with a transition to the loss-dominated regime over the energy range 0.015--60~TeV. The local electron spectra are determined by the injection spectrum of the sources, modified by energy losses during propagation, $\alpha(\mathbf{r},E)$, and by diffusion, $D_{\parallel,\perp}(E) \propto E^\delta$. 

The injection spectrum of primary electrons accelerated in SNRs is 
parameterized as a smoothly broken power law with an additional 
high-energy suppression. Since our calculations start at 15~GeV, above the low-energy break $E_{low} \sim 10$~GeV, the low-energy 
transition is not relevant in this energy range. In practice, in our Monte Carlo and diffusion calculations, we vary the effective 
high-energy slope $\gamma$, which already includes both the 
acceleration index and the additional spectral softening due to 
particle escape from the source environment. Therefore, the SNR 
injection spectrum used in the propagation code can be effectively 
approximated as a single power law with a smooth high-energy 
suppression around $E_c$,
\begin{equation}
\begin{split}
Q_{\rm SNR}(E,r) =
Q_0
&\left(\frac{E}{E_c}\right)^{-\gamma}
\left[
1+
\left(\frac{E}{E_c}\right)^{1/s}
\right]^{-s},
\label{eq:snr_injection}
\end{split}
\end{equation}

where $Q_0$ is the normalization, $s$ controls the smoothness of 
the spectral transition, and $E_c$ determines the onset of the 
high-energy suppression.

In addition to SNRs, we include contributions from electrons and positrons produced in pulsar wind nebulae (PWNe). The PWN injection spectrum is
modeled as a broken power law with an exponential cutoff,

\begin{equation}
\begin{split}
Q_{\rm PWN}(E,r) = \\
Q_{\rm PWN}
\begin{cases}
\left(\dfrac{E}{E_b}\right)^{-\gamma_1}, & E < E_b, \\
\left(\dfrac{E}{E_b}\right)^{-\gamma_2}
\exp\!\left(-\dfrac{E}{E_{\max}}\right), & E \ge E_b ,
\end{cases}
\label{eq:pwn_injection}
\end{split}
\end{equation}

where $\gamma_1$ and $\gamma_2$ are the low- and high-energy slopes,
$E_b$ is the break energy, and $E_{\max}$ is the cutoff energy. Following Ref.~\cite{AmatoBlasi2017}, we adopt 
$\gamma_1 \simeq 1.5$ below $E_b=550$~GeV and $\gamma_2 \simeq 2.4$ above the break. We find that a PWN contribution of $\epsilon_{\rm PWN}=9\%$ relative to the SNR component in the vicinity of the Solar System is sufficient to reproduce the data. However, larger values up to $\epsilon_{\rm PWN}=15\%$ have been reported, e.g., in Ref.~\cite{EvoliBlasiAmatoAloisio2020}, and the overall pulsar contribution to the observed spectrum remains the subject of active discussion~\cite{ManconiDiMauroDonato2020}.

We assume that primary electrons are accelerated at SNR shock fronts~\cite{ZirakashviliAharonian2007}. The spatial distribution of SNRs follows the spiral-arm structure adopted in Ref.~\cite{BlasiAmato2012}. The effective high-energy slope $\gamma$, the normalization $Q_0$, and the break energy in the injection spectrum $E_c$ are treated as free model parameters. We also model the secondary electrons and positrons produced in interactions of CR nuclei with the interstellar medium (ISM). The gas distribution is adopted from Ref.~\cite{Strong1998}, and we treat CR protons as the dominant channel for secondary lepton production. The proton spectrum is computed at every grid point of the Galactic volume so as to reproduce the AMS-02 spectrum at the Solar position, in agreement with our earlier work~\cite{borisov2025}. We find that the secondary contribution remains subdominant across the entire energy range, reaching at most $\sim 4\%$ of the primary SNR electron flux at 50~GeV.

\begin{figure*}
\includegraphics[width=0.95\textwidth]{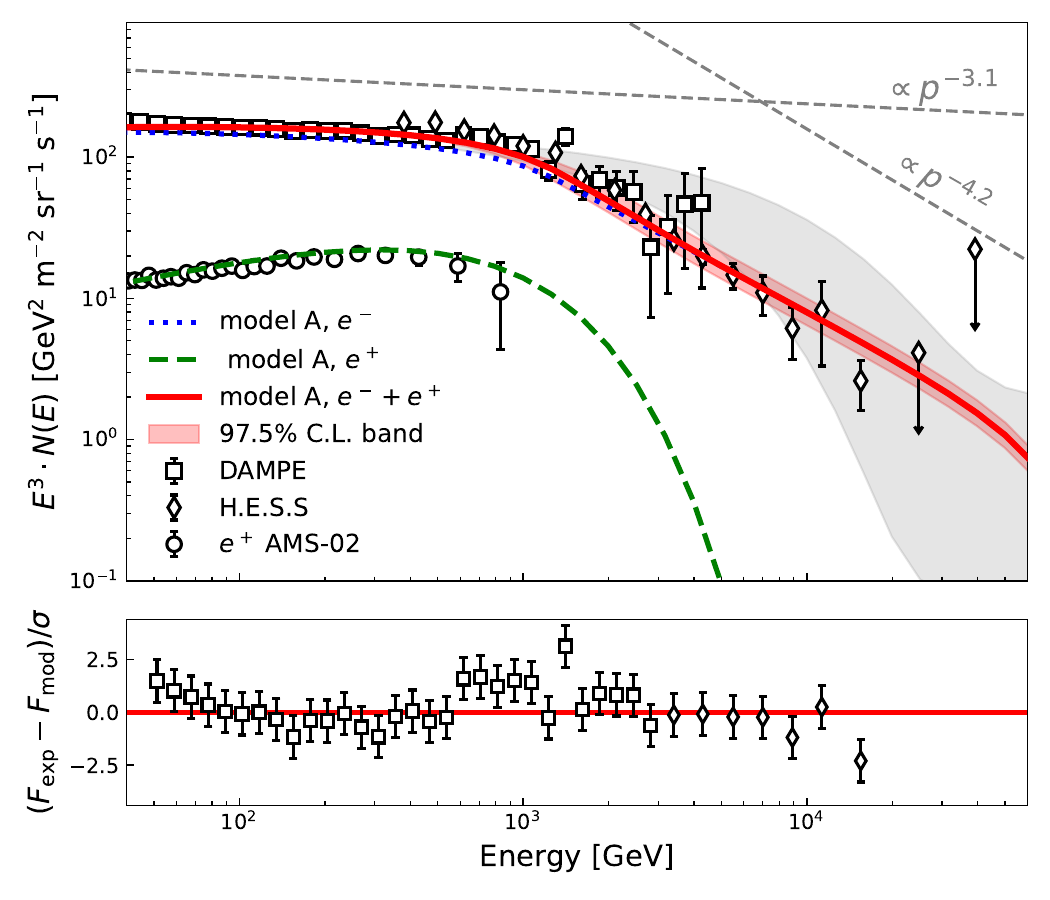}
\caption{\label{fig:e_spectrum}
Fit to the combined $e^-+e^+$ spectrum using DAMPE~\cite{dampe2017}, H.E.S.S.\cite{hess2024} data for the total lepton flux and AMS-02~\cite{ams02electrons2014} data for the positron fraction over the energy interval 0.04--60~TeV. The upper panel shows the best-fit spectrum, and the lower panel shows the residuals normalized to $\sigma$. DAMPE data points overlapping with the H.E.S.S.\ energy range, as well as H.E.S.S.\ points with only upper energy bounds, were excluded from the fit. The shaded gray band represents the uncertainty associated with the stochastic distribution of nearby sources, following Ref.~\cite{EvoliAmatoBlasiAloisio2020Stochastic}.
}
\end{figure*}

Figure~\ref{fig:e_spectrum} shows the best-fit spectrum corresponding 
to the optimal parameter sets $(\gamma,Q_0,E_c)$ 
obtained in the Monte Carlo exploration of the anisotropic model. 
In our calculations, we treat $\gamma$ as an effective, 
free parameter representing the combined effect of acceleration at 
the SNR shock and the additional softening due to particle escape 
from the source environment. Reference values in the literature 
tend to be slightly higher than the canonical acceleration index 
(e.g., $\gamma_0 \simeq 2$) because they include this escape-induced softening. Because of the computational cost, the minimization was performed 
in two stages: first on a coarse grid and then on a refined grid 
around the lowest local minimum for the parameter ranges listed 
in Table~\ref{tab:table1}.

\begin{table}[b]
\caption{\label{tab:table1}
Number of realizations $N$, model parameters ($\gamma$, $Q_0$, $E_c$), and parameter ranges explored in the fitting procedure for Models A and B.}
\begin{ruledtabular}
\begin{tabular}{cccc}
 N &$\gamma$& $Q_0$ & $E_c$ \\
 &  & (TeV$^{-1}$ s$^{-1}$ kpc$^{-3}$)& (TeV) \\
\hline
\multicolumn{4}{c}{Model A} \\
$3375$ & 2.100--2.250 & $6\times 10^{-23}$--$1\times10^{-21}$ & 0.8--4.0 \\
$343$  & 2.169--2.171 & $4\times 10^{-22}$--$4\times10^{-22}$ & 1.2--1.4 \\
\multicolumn{4}{c}{Model B} \\
$3375$ & 2.200--2.400 & $5\times10^{-22}$--$1\times10^{-20}$ & 0.8--4.0 \\
$343$  & 2.342--2.345 & $9\times10^{-21}$--$9\times10^{-21}$ & 1.1--1.3 \\
\end{tabular}
\end{ruledtabular}
\end{table}

We perform calculations for two models. 

Model~A includes both the spatial and energy dependence of the full diffusion tensor $\hat{D}_{ij}(\mathbf{r},E)$ as described in Sec.~\ref{sec:tensor}, i.e., the fully anisotropic case. The parallel and perpendicular diffusion coefficients are computed at each grid point using a trajectory method for a fixed turbulence correlation length ($\ell_c = 50$ pc) and the local ratio of regular to turbulent magnetic field, with the regular field modeled according to Ref.~\cite{Unger_2024}. The resulting diffusion tensor is then rotated into Galactic Cartesian coordinates.

Model~B assumes spatially uniform diffusion coefficients, $D_{\parallel} = D_{\perp} = D_0$, with a simple power-law scaling $D(E) = D_0 \left(\frac{E}{E_0}\right)^{1/3}$, i.e., the isotropic case. The normalization $D_0 = 4.5\times 10^{28}$ cm$^2$/s, $E_0 = 4$ GeV, and spectral index are chosen to reproduce the local cosmic-ray spectrum~\cite{Delahaye2010}.

Model~B is not shown in Fig.~\ref{fig:e_spectrum} because its spectrum, for the best-fit realization, remains within the uncertainty band of Model~A and reproduces the local spectral shape comparably well.

\begin{table}[b]
\caption{\label{tab:table2}
Best-fit values of the parameters ($\gamma$, $Q_0$, $E_c$) and the $\chi^2$ values of the fits.}
\begin{ruledtabular}
\begin{tabular}{ccccc}
 Model & $\gamma$ & $Q_0$ & $E_c$ & $\chi^2$ \\
 &  & (TeV$^{-1}$ s$^{-1}$ kpc$^{-3}$)  & (TeV) & (ndf = 37)\\
\hline
A & 2.169 & $3.169\times10^{-22}$ & 1.333 & 27.18 \\
B & 2.344 & $8.833\times10^{-21}$ & 1.267 & 31.32 \\
\end{tabular}
\end{ruledtabular}
\end{table}

We find that both models can reproduce the observed local $e^-+e^+$ spectrum; however, the isotropic Model~B requires a significantly larger injection index than the anisotropic Model~A, in which the spectrum naturally softens due to the enhanced contribution of sources preferentially located along the direction of the regular Galactic magnetic field. We find that the observed lepton spectrum can be reproduced by the background SNR distribution for the primary electrons, together with the secondary component. 

The calculations presented in this work adopt the standard continuous source approximation. However, the local electron spectrum at high energies may be influenced by the discrete and stochastic nature of the source population. A quantitative treatment of this effect requires modeling individual sources, which is typically modeled using Green's function approaches combined with Monte Carlo realizations of the Galactic source population. Since our diffusion solver assumes a continuous source distribution, we do not model this effect self-consistently. Instead, we illustrate its expected magnitude using the uncertainty band derived from previous studies of stochastic electron sources. In particular, the band shown in Fig.~\ref{fig:e_spectrum} is based on the results of Ref.~\cite{EvoliAmatoBlasiAloisio2020Stochastic}, where the spread of the local spectrum was estimated for the isotropic case using a Green's function approach. These estimates rely on simplified transport models and should therefore be interpreted only as an order-of-magnitude indication of the possible spread of the local spectrum at TeV energies. Below $\sim 1$~TeV, the stochastic uncertainty remains relatively small, and the difference between Model~A and Model~B primarily reflects the different transport properties of the two diffusion scenarios. In this regime, the anisotropic diffusion model requires a harder injection spectrum than the isotropic approximation. At higher energies, the stochastic contribution becomes significant and may partially obscure the differences between the two transport models. Nevertheless, the need for different injection indices arises from the distinct propagation properties of the models rather than from the stochasticity of the sources.

\begin{figure*}
\includegraphics[width=0.95\textwidth]{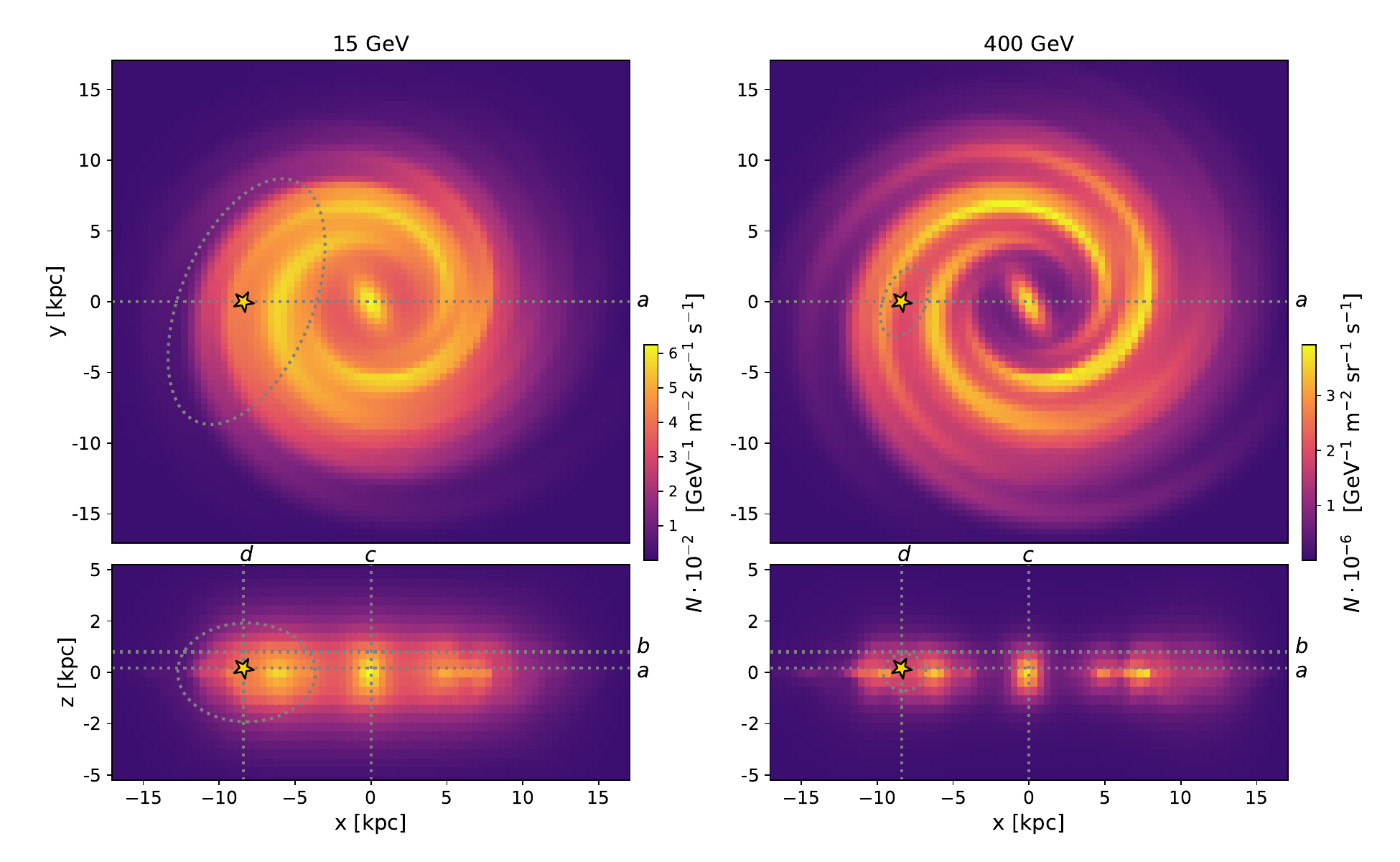}
\caption{\label{fig:dist}
Spatial distribution of CR leptons in Model~A (anisotropic diffusion) at 15~GeV (left) and 400~GeV (right). The model parameters correspond to the best fit in Table~\ref{tab:table2}. The Solar position is marked with a star at $(x,y,z)=(-8.2,0.0,0.2)$~kpc. The ellipses indicated by gray dashed lines represent the approximate shape of the lepton propagation horizon.
}
\end{figure*}

For the best-fit parameters of Model~A ($\gamma = 2.169$, $Q_0 = 3.169\times10^{-22}$ TeV$^{-1}$ s$^{-1}$ kpc$^{-3}$ and $E_c = 1.333$ TeV), Fig.~\ref{fig:dist} shows the $e^-+e^+$ density in two Cartesian slices of the Galaxy.

The left panel corresponds to 15~GeV, where diffusion and energy losses contribute comparably to the spectral shape. The right panel corresponds to 400~GeV, where losses dominate over diffusion, strongly limiting the distance from which CR leptons can reach the observer. The upper panels show a slice at $z=0.2$~kpc through the Solar position, and the lower panels show a slice at $y=0.0$~kpc. The Solar position is indicated by a star.

The ellipses indicated by gray dashed lines mark the approximate lepton propagation horizon in the anisotropic case (Model~A).
Following the formalism of Refs.~\cite{EvoliBlasiAmatoAloisio2020} and~\cite{BressloffLawley2015} for the spatial dependence of the diffusion-tensor components, we compute the horizon as the characteristic distance from which leptons of a given energy can reach the observer along the principal directions of the local magnetic-field frame, rather than assuming equal diffusion radii as in isotropic models.
This yields
\begin{equation}
\begin{split}
\lambda_i^2(E,E_s) = 4 \int_{E}^{E_s} \frac{\langle D_i^{\rm fld}(\mathbf{x},E')\rangle}{\alpha(\mathbf{x},E')} \, dE' \; ,\\ \quad i=1,2,3,
\label{eq:lambda_general}
\end{split}
\end{equation}
where $\langle D_i^{\rm fld}(\mathbf{x},E')\rangle$ denotes the $i$th component of the diffusion tensor in the local magnetic-field frame, given by Eq.~(\ref{eq:Drot}). Here, $i=1$ corresponds to the direction along the regular magnetic field, while $i=2,3$ denote the perpendicular directions. The true propagation region is not strictly ellipsoidal but has a more complex geometry, elongated along the regular field lines and reflecting the spatial dependence of both the diffusion tensor and the loss function, qualitatively consistent with Fig.~\ref{fig:dist}.

\begin{figure*}
\includegraphics[width=0.95\textwidth]{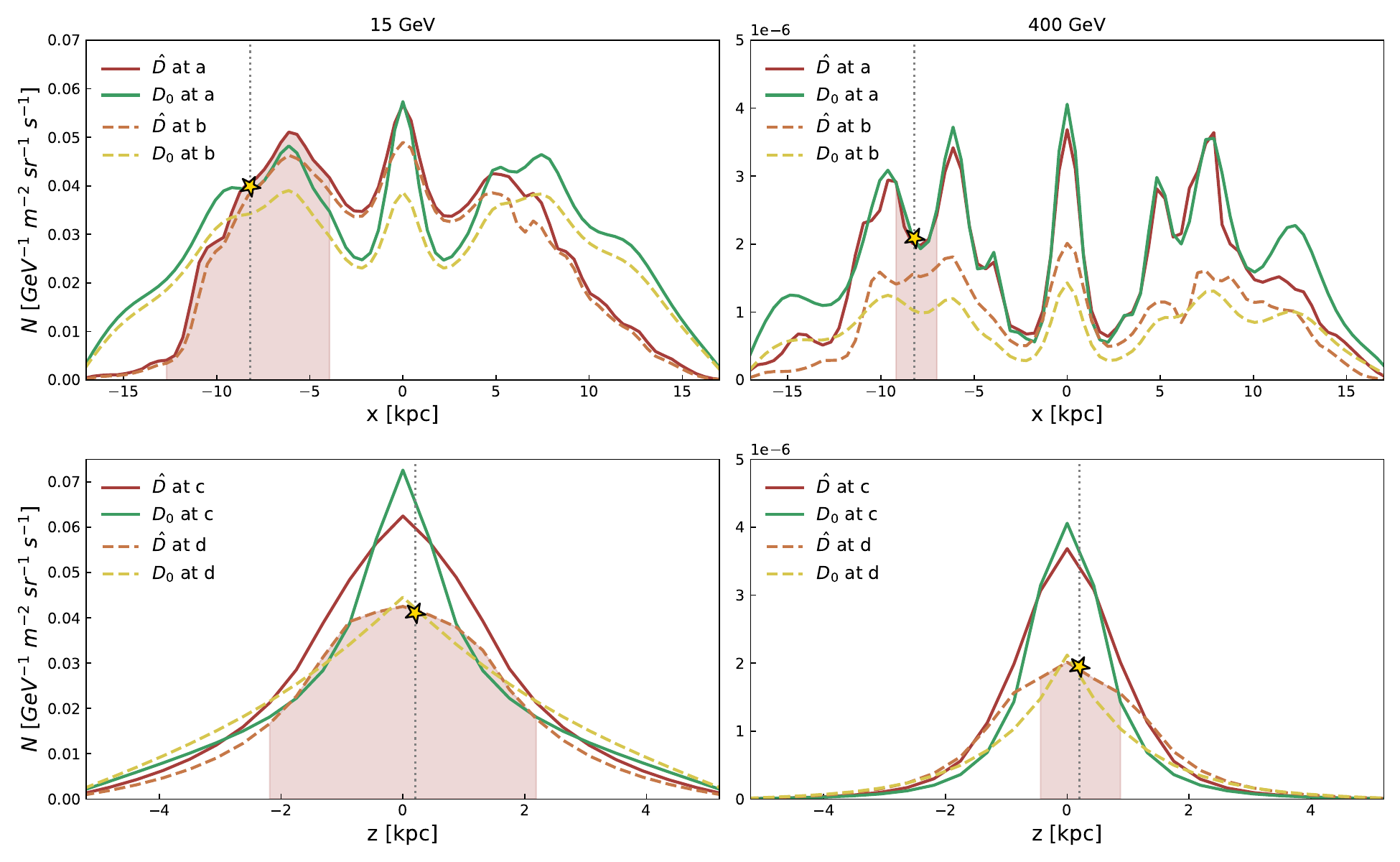}
\caption{\label{fig:dist_c}
Comparison of $e^-+e^+$ distributions in Model~A (anisotropic diffusion $\hat{D}$) and Model~B (isotropic diffusion $D_0$) at 15~(left) and 400~GeV (right). The profiles are shown along $x$ at locations a $(y=0.0,z=0.2)$ and b $(y=0.0,z=1.0)$ and along $z$ at locations c $(x=0.0,y=0.0)$ and d $(x=-8.2,y=0.0)$~kpc. The shaded regions indicate the characteristic lepton horizon. The Solar position is marked with a star at $(x,y,z)=(-8.2,0.0,0.2)$~kpc.
}
\end{figure*}

For 15~GeV, we obtain $\lambda_{1,2,3}=\{9.1,\,4.4,\,3.8\}$~kpc, decreasing to $\{1.3,\,0.6,\,0.5\}$~kpc at 400~GeV, where $\lambda_{1,2,3}$ denote the principal semi-axes of the effective propagation ellipsoid in the local magnetic-field frame. Figure~\ref{fig:dist_c} shows a comparison of the propagated fluxes in Models~A and B along the $x$ and $z$ directions for four positions in the Galaxy, marked as a, b, c and d in Fig.~\ref{fig:dist}.

Both models reproduce the local lepton density near the Solar position, but at 15~GeV they show significant differences along the $x$ direction, particularly in the interarm regions and near the Galactic edges. Model~A yields a narrower and asymmetric profile, reflecting the preferential transport of CR leptons along the regular magnetic-field direction, which facilitates CR leakage. The vertical ($z$) profiles in the anisotropic Model~A are broader than in Model~B due to enhanced leakage into the Galactic halo. At 400~GeV, the anisotropic and isotropic profiles become more similar, reflecting the reduced lepton horizon and the increasing dominance of the source distribution over transport effects. Significant differences persist only in regions far from the sources—at large Galactocentric radii and above the disk—where anisotropic leakage continues to play a significant role.
 
\section{\label{sec:conclusion}Conclusion}
We have presented a method for constructing a fully spatially dependent anisotropic diffusion tensor based on direct numerical integration of charged-particle trajectories in a realistic Galactic magnetic field. We incorporate this tensor into a finite-difference transport scheme that explicitly accounts for spatial gradients of its components. This approach provides a self-consistent framework for modeling cosmic-ray propagation in which spatial variations of the diffusion tensor affect the resulting lepton distribution.

We have shown that the observed local spectrum of cosmic-ray electrons and positrons can be reproduced within a diffusion framework operating in the transition toward the loss-dominated regime, at energies of a few tens of GeV. In this regime the spectrum can be described without introducing multiple breaks either in the injection spectrum or in the energy dependence of the diffusion tensor. Under mixed transport conditions, anisotropic diffusion produces a narrower spatial distribution of cosmic-ray leptons due to preferential propagation along magnetic-field lines. In the loss-dominated regime, energy losses restrict the effective propagation distance of leptons, making the spatial distribution of sources an important factor shaping the observed spectrum.

We find that allowing for spatially varying anisotropic diffusion leads to a systematically harder injection spectrum compared to isotropic diffusion models. This difference persists over the entire energy range considered. In the anisotropic scenario the spatial dependence of the diffusion tensor and preferential transport along the regular magnetic field introduce an additional spectral steepening during propagation. As a result, the observed local lepton spectrum can be reproduced with a harder injected spectrum than required in the isotropic approximation. 

At the highest energies, the local electron spectrum can be affected by the stochastic distribution of nearby sources, which can introduce significant fluctuations in the lepton flux, particularly above the TeV scale where radiative losses restrict the effective propagation distance. These fluctuations contribute an additional uncertainty in the interpretation of the observed spectrum. Within the present approach, the injection spectrum is determined through Monte Carlo simulations over a broad energy range (15~GeV--60~TeV) without introducing additional spectral breaks, apart from the standard steepening associated with the source physics around $\sim$1~TeV. The difference in the required injection spectra between the anisotropic and isotropic diffusion models therefore primarily reflects the distinct transport properties, with the anisotropic scenario yielding injection indices that are closer to theoretical expectations. A fully self-consistent treatment of stochastic source populations within anisotropic diffusion models remains an important subject for future work.

\begin{acknowledgments}
The research was supported by RSF (Project No.25-22-00246).
\end{acknowledgments}

\section*{DATA AVAILABILITY}
The experimental data used in this work are publicly available from Refs.~\cite{dampe2017}, ~\cite{hess2024}, and ~\cite{ams02electrons2014}. For a more detailed description of the mathematical framework employed in this study,
see Refs.~\cite{borisov2025}, ~\cite{borisov2025_2}, and ~\cite{BorisovKudryashov2025}. The code developed
for data processing and analysis is available from the corresponding author upon reasonable request.


\begin{thebibliography}{}
\end{thebibliography}

\begin{thebibliography}{99}

\bibitem{Hutschenreuter_2024}
S. Hutschenreuter, M. Haverkorn, P. Frank et al., Disentangling the Faraday rotation sky, 
\href{https://doi.org/10.1051/0004-6361/202346740}{\em Astron. Astrophys. 690 A314 (2024)}.

\bibitem{Heinrich_2024}
A. Tsouros, G. Edenhofer, T. Enßlin et al., Reconstructing Galactic magnetic fields from local measurements for backtracking ultra-high-energy cosmic rays, \href{https://doi.org/10.1051/0004-6361/202346423}{\em Astron. Astrophys. 681, A111 (2024)}.

\bibitem{Prevotat_2024}
C. Prevotat, M. Kachelrieß, S. Koldobskiy,
A. Neronov, and D. Semikoz, Energy dependence of the knee in the cosmic-ray spectrum across the Milky Way, 
\href{https://doi.org/10.1103/PhysRevD.110.103035}{{\em Physical Review D} 110(10):103035 (2024)}.


\bibitem{StrongMoskalenkoReimer2004}
A. W. Strong, I. V. Moskalenko, and O. Reimer,
Diffuse Galactic continuum gamma rays. A model compatible with EGRET data and cosmic-ray measurements,
\href{https://doi.org/10.1086/423193}
{{\em Astrophys. J.} 613, 962 (2004)}.

\bibitem{Gaggero2015DRAGON}
D. Gaggero, D. Grasso, A. Marinelli et al., The gamma-ray and neutrino sky: A consistent picture of Fermi-LAT, Milagro, and IceCube results,
\href{https://doi.org/10.1088/2041-8205/815/2/L25}{\em Astrophys. J. Lett. 815, (2015)}.

\bibitem{Kissmann2014PICARD}
R. Kissmann, PICARD: A novel code for the Galactic Cosmic Ray propagation problem,
\href{https://doi.org/10.1016/j.astropartphys.2014.02.002}{{\em Astrophys. J.}55, 37 (2014)}.

\bibitem{Ackermann2012Diffuse}
M. Ackermann et al. (Fermi-LAT Collaboration), Fermi-LAT observations of the diffuse gamma-ray emission: implications for cosmic rays and the interstellar medium,\href{https://doi.org/10.1088/0004-637X/750/1/3}{{\em Astrophys. J.} 750, 3 (2012)}.

\bibitem{OrlandoStrong2013}
E. Orlando and A. W. Strong,
Galactic synchrotron emission with cosmic ray propagation models,
\href{https://doi.org/10.1093/mnras/stt1718}
{{\em Mon. Not. R. Astron. Soc.} 436, 2127 (2013)}.

\bibitem{Webb_1994}
G. M. Webb, J. R. Jokipii, and G. E. Morfill, Green's formula and variational principles for cosmic-ray transport with application to rotating and shearing flows, \href{https://doi.org/10.1086/173880}{{\em Astrophys. J.} 424, 158 (1994)}.

\bibitem{Hanasz_2021}
M. Hanasz, A. Strong, P. Girichidis, Simulations of cosmic ray propagation, \href{https://doi.org/10.1007/s41115-021-00011-1}{{\em Living Rev. Comput. Astrophys.} 7, 2 (2021)}.

\bibitem{Strong1998}
A. W. Strong and I. V. Moskalenko, Propagation of cosmic-ray nucleons in the galaxy, \href{https://doi.org/10.1086/306470}{{\em Astrophys. J.} 509, 212 (1998)}.

\bibitem{Evoli_2017}
C. Evoli, D. Gaggero, A. Vittino, G. Di Bernardo, M. Di Mauro, A. Ligorini, P. Ullio, and D. Grasso, Cosmic-ray propagation with DRAGON2: I. numerical solver and astrophysical ingredients, \href{https://doi.org/10.1088/1475-7516/2017/02/015}{{\em JCAP} 02, 015 (2017)}.

\bibitem{Batista_2016}
R. A. Batista, A. Dundovic, M. Erdmann, K.-H. Kampert, D. Kuempel, G. Müller, G. Sigl, A. van Vliet, D. Walz, and T. Winchen, CRPropa 3 — a public astrophysical simulation framework for propagating extraterrestrial ultra-high energy particles, \href{https://doi.org/10.1088/1475-7516/2016/05/038}{{\em JCAP} 05, 038 (2016)}.

\bibitem{Giacinti_2012}
G. Giacinti, M. Kachelriess, and D. V. Semikoz, Filamentary diffusion of cosmic rays on small scales, \href{https://doi.org/10.1103/PhysRevLett.108.261101}{{\em Phys. Rev. Lett.} 108, 261101 (2012)}.

\bibitem{Prosekin_2015}
A. Y. Prosekin, S. R. Kelner, and F. A. Aharonian, On transition of propagation of relativistic particles from the ballistic to the diffusion regime, \href{https://doi.org/10.1103/PhysRevD.92.083003}{{\em Phys. Rev. D} 92, 083003 (2015)}.

\bibitem{kuhlen_2025}
M. Kuhlen, V. H. M. Phan, and P. Mertsch, Diffusion of relativistic charged particles and field lines in isotropic turbulence: I. Numerical simulations, \href{https://doi.org/10.3847/1538-4357/adee9a}{{\em Astrophys. J.} 992, 10 (2025)}.

\bibitem{Wibking_2023}
B. D. Wibking, The global structure of magnetic fields and gas in simulated Milky Way-analogue galaxies, 
\href{https://doi.org/10.1093/mnras/stac2648}{{\em Mon. Not. Roy. Astron. Soc.} 521, 5972 (2023)}.

\bibitem{Giacinti_2015}
G. Giacinti, M. Kachelrieß, and D. V. Semikoz, Escape model for Galactic cosmic rays and an early extragalactic transition, \href{https://doi.org/10.1103/PhysRevD.91.083009}{{\em Phys. Rev. D} 91, 083009 (2015)}.

\bibitem{Jansson_2012}
R. Jansson and G. R. Farrar, A new model of the Galactic magnetic field, \href{https://doi.org/10.1088/0004-637X/757/1/14}{{\em Astrophys. J.} 757, 14 (2012)}.

\bibitem{Unger_2024}
M. Unger and G. R. Farrar, The coherent magnetic field of the Milky Way, \href{https://doi.org/10.3847/1538-4357/ad4a54}{{\em Astrophys. J.} 970, 95 (2024)}.

\bibitem{Casse_2002}
R. Casse, M. Lemoine, and G. Pelletier, Transport of cosmic rays in chaotic magnetic fields, \href{https://doi.org/10.1103/PhysRevD.65.023002}{{\em Phys. Rev. D} 65, 023002 (2002)}.

\bibitem{Dundovic_2020}
A. Dundovic, O. Pezzi, P. Blasi, C. Evoli, and W. H. Matthaeus, Novel aspects of cosmic ray diffusion in synthetic magnetic turbulence, \href{https://doi.org/10.1103/PhysRevD.102.103016}{{\em Phys. Rev. D} 102, 103016 (2020)}.


\bibitem{Shalchi_2009}
M. Shalchi, Nonlinear cosmic ray diffusion theories, (Springer, Berlin, 2009).

\bibitem{DeLaTorreLuque_2025}
A. Ambrosone, C. Evoli, B. Schroer et al., The origin of very high-energy diffuse $\gamma$-ray emission: The case for galactic source cocoons,
\href{https://doi.org/10.1051/0004-6361/202554796}
{{\em Astron. Astrophys.} 698, L18 (2025)   }.


\bibitem{ALZetoun2018Anisotropic}
A. Al‐Zetoun and A. Achterberg,
Propagation of Galactic cosmic rays: the influence of anisotropic diffusion,
\href{https://doi.org/10.1093/mnras/sty727}
{{\em Mon. Not. R. Astron. Soc.} 477, 1258 (2018)}.

\bibitem{dampe2017}
DAMPE Collaboration, Direct detection of a break in the teraelectronvolt cosmic‑ray spectrum of electrons and positrons, \href{https://doi.org/10.1038/nature24475}{{\em Nature} 552, 63-66 (2017)}.

\bibitem{adriani2017}
O. Adriani et al. (the CALET Collaboration), Energy spectrum of cosmic-ray electron and positron from 10 GeV to 3 TeV observed with the calorimetric electron telescope on the International Space Station, \href{https://doi.org/10.1103/PhysRevLett.119.181101}{{\em Phys. Rev. Lett.} 119, 181101 (2017)}.

\bibitem{fermilat2017}
S. Abdollahi, M. Ackermann, M. Ajello, et al. (the Fermi-LAT Collaboration), Cosmic-ray electron-positron spectrum from 7 GeV to 2 TeV with the Fermi Large Area Telescope, \href{https://doi.org/10.1103/PhysRevD.95.082007}{{\em Phys. Rev. D} 95, 082007 (2017)}.

\bibitem{hess2024}
F. Aharonian, F. Ait Benkhali, J. Aschersleben, et al. (the H.E.S.S. Collaboration), High-statistics measurement of the cosmic-ray electron spectrum with H.E.S.S., \href{https://doi.org/10.1103/PhysRevLett.133.221001}{{\em Phys. Rev. Lett.} 133, 221001 (2024)}.

\bibitem{ams02electrons2014}
M. Aguilar, L. Ali Cavasonza, G. Ambrosi, et al. (the AMS-02 Collaboration), Precision measurement of the electron flux in primary cosmic rays from 0.5 GeV to 1 TeV with AMS-02, \href{https://doi.org/10.1103/PhysRevLett.113.221102}{{\em Phys. Rev. Lett.} 113, 221102 (2014)}.

\bibitem{EvoliBlasiAmatoAloisio2020}
C. Evoli, P. Blasi, E. Amato, and R. Aloisio, The signature of energy losses on the cosmic ray electron spectrum, \href{https://doi.org/10.1103/PhysRevLett.125.051101}{{\em Phys. Rev. Lett.} 125, 051101 (2020)}.

\bibitem{thoudam2024}
S. Thoudam, Origin of the break in the cosmic-ray electron plus positron spectrum at ~1 TeV, \href{https://doi.org/10.1051/0004-6361/202348607}{{\em Astron. Astrophys.} 690, A351 (2024)}.

\bibitem{dellatorreluque2021}
P. de la Torre Luque, M. N. Mazziotta, F. Loparco, F. Gargano, and D. Serini, Markov chain Monte Carlo analyses of the flux ratios of B, Be and Li with the DRAGON2 code, \href{https://doi.org/10.1088/1475-7516/2021/07/010}{{\em JCAP} 07, 010 (2021)}.

\bibitem{niu2018}
J.-S. Niu and T. Li, Galactic cosmic-ray model in the light of AMS-02 nuclei data, \href{https://doi.org/10.1103/PhysRevD.97.023015}{{\em Phys. Rev. D} 97, 023015 (2018)}.

\bibitem{borisov2025}
V. D. Borisov, V. O. Yurovsky, and A. I. Peryatinskaya, Spatial dependence of the break in the energy spectrum of cosmic rays in the new anisotropic diffusion approach, \href{https://doi.org/10.1103/7572-l2zp}{{\em Phys. Rev. D} 112, 023010 (2025)}.

\bibitem{borisov2025_2}
V. D. Borisov, V. O. Yurovsky, and I. A. Kudryashov, Modulation of the galactic cosmic ray spectrum in an anisotropic diffusion approach, \href{https://doi.org/10.1134/S1062873825711602}{{\em Bull. Russ. Acad. Sci. Phys.} 89, 1019-1023 (2025)}.

\bibitem{Iffrig2017}
O. Iffrig and P. Hennebelle, Structure distribution and turbulence in self-consistently supernova-driven ISM of multiphase magnetized galactic discs, \href{https://doi.org/10.1051/0004-6361/201630290}{{\em Astron. Astrophys.} 604, A70 (2017)}.

\bibitem{Moderski2005ApJ627}
 R. Moderski, M. Sikora, P. Coppi et al., Klein-Nishina effects in the spectra of non-thermal sources immersed in external radiation fields, \href{https://doi.org/10.1111/j.1365-2966.2005.09494.x}{{\em Mon. Not. R. Astron. Soc.} 363, 3 (2005)}.

\bibitem{Delahaye2010}
T. Delahaye, J. Lavalle, R. Lineros, F. Donato, and N. Fornengo, Galactic electrons and positrons at the Earth: new estimate of the primary and secondary fluxes, \href{https://doi.org/10.1051/0004-6361/201014225}{{\em Astron. Astrophys.} 524, A51 (2010)}.

\bibitem{BlumenthalGould1970}
G. R. Blumenthal and R. J. Gould, Bremsstrahlung, synchrotron radiation, and Compton scattering of high-energy electrons traversing dilute gases, \href{https://doi.org/10.1103/RevModPhys.42.237}{{\em Rev. Mod. Phys.} 42, 237-270 (1970)}.

\bibitem{AmatoBlasi2017}
E. Amato and P. Blasi, Cosmic ray transport in the Galaxy: a review, \href{https://doi.org/10.48550/arXiv.1704.05696}{{\em Adv.Space Res.} 62, 2731-2749 (2018)}.

\bibitem{ManconiDiMauroDonato2020}
S. Manconi, M. Di Mauro, and F. Donato, Contribution of pulsars to cosmic-ray positrons in light of recent observation of inverse-Compton halos, \href{https://doi.org/10.1103/PhysRevD.102.023015}{{\em Phys. Rev. D} 102, 023015 (2020)}.


\bibitem{ZirakashviliAharonian2007}
V. N. Zirakashvili and F. Aharonian, Analytical solutions for energy spectra of electrons accelerated by nonrelativistic shock-waves in shell type supernova remnants, \href{https://doi.org/10.1051/0004-6361:20066494}{{\em Astron. Astrophys.} 465, 695-702 (2007)}.

\bibitem{BlasiAmato2012}
P. Blasi and E. Amato, Diffusive propagation of cosmic rays from supernova remnants in the Galaxy. I: spectrum and chemical composition, \href{https://doi.org/10.1088/1475-7516/2012/01/010}{{\em JCAP} 01, 010 (2012)}.

\bibitem{EvoliAmatoBlasiAloisio2020Stochastic}
C. Evoli, E. Amato, P. Blasi et al, Stochastic nature of Galactic cosmic-ray sources,
\href{https://doi.org/10.1103/PhysRevD.104.123029}
{{\em Phys. Rev. D} 104, 123029 (2021)}.

\bibitem{BressloffLawley2015}
P. C. Bressloff and S. D. Lawley, Moment equations for a piecewise deterministic PDE, \href{https://doi.org/10.1088/1751-8113/48/10/105001}{{\em J. Phys. A: Math. Theor.} 48, 105001 (2015)}.

\bibitem{BorisovKudryashov2025}
V. D. Borisov and I. A. Kudryashov, Four-Dimensional Anisotropic Diffusion of Cosmic Rays in Realistic Galactic Magnetic Fields, \href{https://doi.org/10.22323/1.501.0016}{{\em PoS(ICRC2025)} 016 (2025)}.


\end{thebibliography}
\end{document}